\documentstyle[11pt,newpasp,twoside,epsf]{article}
\def\fb{{178 {\rm \thinspace MHz}}}

\def\edcomment#1{\iffalse\marginpar{\raggedright\sl#1\/}\else\relax\fi}
\reversemarginpar

\begin{document}
 \title{Radio Galaxy Spectra}
 \author{C A Jackson}
\affil{Research School of Astronomy \& Astrophysics, 
The Australian National University, Cotter Road, Weston Creek,
ACT 2611, Australia}
\author{J V Wall}
\affil{Department of Astrophysics, University of Oxford, Nuclear and
 Astrophysics Laboratory, Keble Road, Oxford OX1 3RH, UK}

\begin{abstract}
Radio spectra of radio galaxies are
often ascribed a simple power law form ($S_{\nu} \propto \nu^{\alpha}$)
with the spectral index, $\alpha$, being of order $-0.7$. 
However, all radio galaxies deviate from this simple 
power law behaviour.
In this paper we derive simple expressions for the average rest-frame
spectra of FRI and FRII radio galaxies. 
These will be used to describe the spectral curvature 
of the parent (FRI and FRII)
populations in models of radio source evolution.
\end{abstract}

The deviation of radio galaxy spectra from simple power law behaviour
occurs at two levels: {\it (i)} the rest-frame spectral index at low
frequency ($<$ 750 MHz) depends on the luminosity of the source -- the
$P - \alpha$ correlation (Blundell, Rawlings \& Willott 1999) and {\it
(ii)} the highest-power sources exhibit the strongest spectral
curvature at high radio frequencies ($>$ 1 GHz) (Laing \& Peacock
1980).  In order to pursue models of radio-source evolution, we wish
to find simple expressions for the average rest-frame spectrum as a
function of radio power and Fanaroff Riley type (Fanaroff \& Riley
1974) over a wide radio frequency range.

The 3CRR sample provides a complete
set of radio galaxies at $S_{178 \rm \thinspace MHz} \ge$ 10.9
Jy (Laing, Riley \& Longair 1983).
We have collected flux-density data for 21 FRI and 61 FRII 
3CRR radio galaxies from measurements
made between 10 and 14.9 GHz.  
Of the FRIIs, we selected the narrow-line radio galaxies only,
to avoid any effects due to orientation. In addition we
selected only sources at $z <$ 0.3) to avoid incorporating
spectral index -- age effects. We scaled 
the individual survey flux density measurements according to the 
factors given in Laing \& Peacock (1980).

\begin{figure}
% from sedfitx.f
\plotfiddle{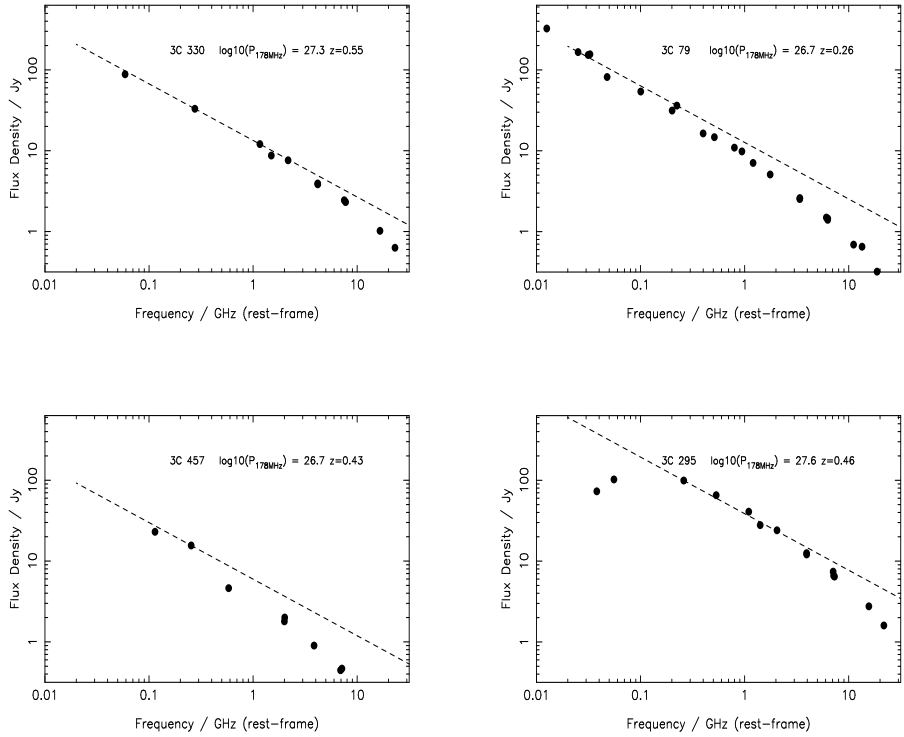}{3.0in}{270}{44}{52}{-175}{270}
\vspace*{0.45in}\caption{
Example spectra for 4 FRII radio galaxies. Observed 
data points ($\bullet$). 
Dashed line indicates slope of $\alpha=-0.7$}
\end{figure}

Typical examples of the spectra of our selected radio galaxies are
shown in Figure 2. It can be seen that some are close to having
$\alpha=-0.7$ over the freqency range shown ({\it i.e.} 3C 330),
others are flatter than $-0.7$, particularly at $<$ 1 GHz ({\it i.e.}
3C 295), while still others steepen above $\sim$1 GHz ({\it i.e.} 3C
79 and 3C 457).

\begin{figure}
% from sedfit1a.f
\plotfiddle{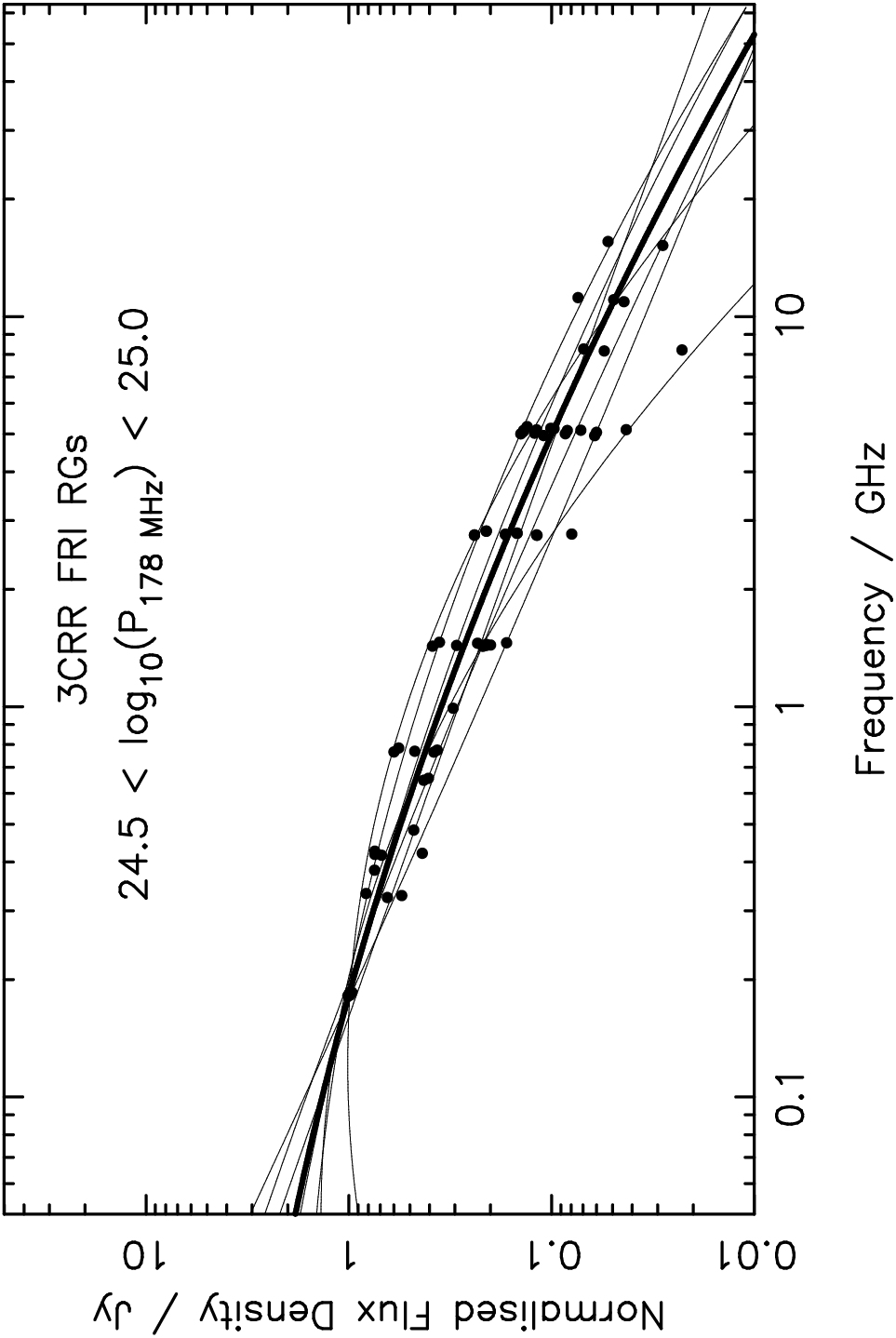}{3.0in}{270}{25}{35}{-190}{300}
\plotfiddle{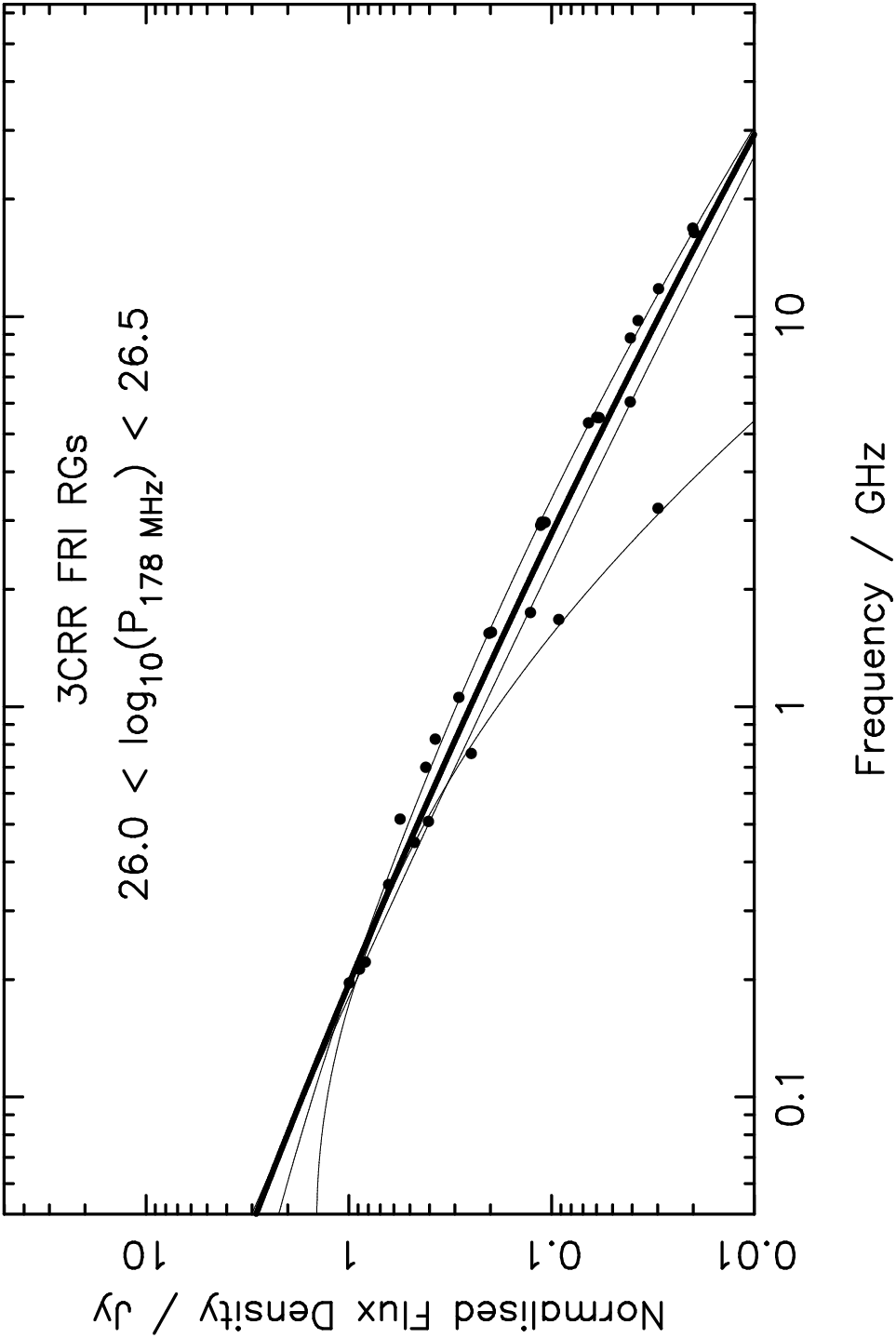}{3.0in}{270}{25}{35}{0}{530}
\vspace*{-4.5in}\caption{Fitted
spectra for individual FRIs (light lines) with radio powers
in the ranges show. The data are
($\bullet$) normalised to $S_{\fb}$ = 1 Jy. The
mean spectrum is the best-fit fo
$\nu_{rest} \ge$ 178 MHz (heavy line).}
\end{figure}

\begin{figure}
% from sedfital.f
\vspace*{0.3in}
\plotfiddle{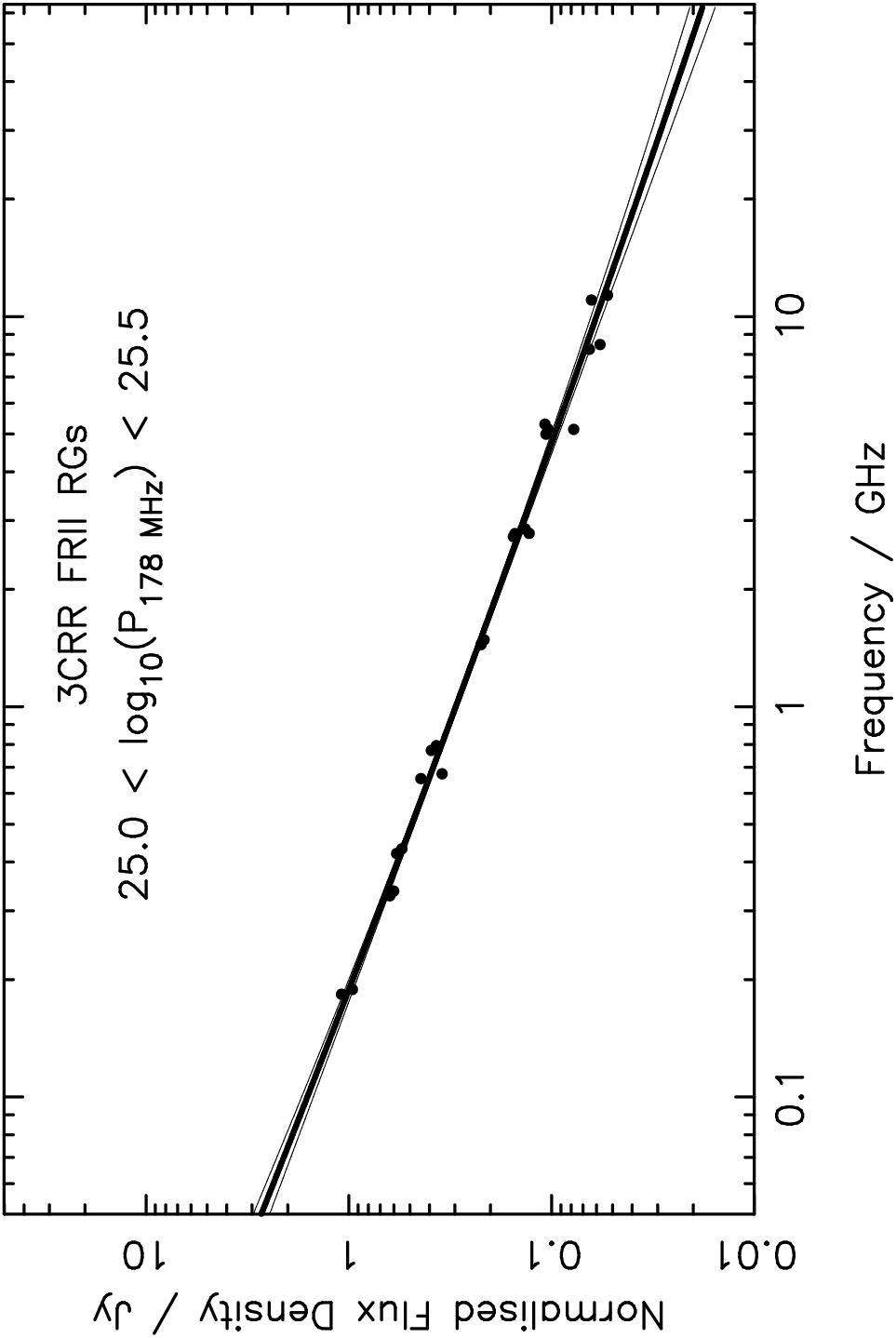}{3.0in}{270}{25}{35}{-180}{270}
\plotfiddle{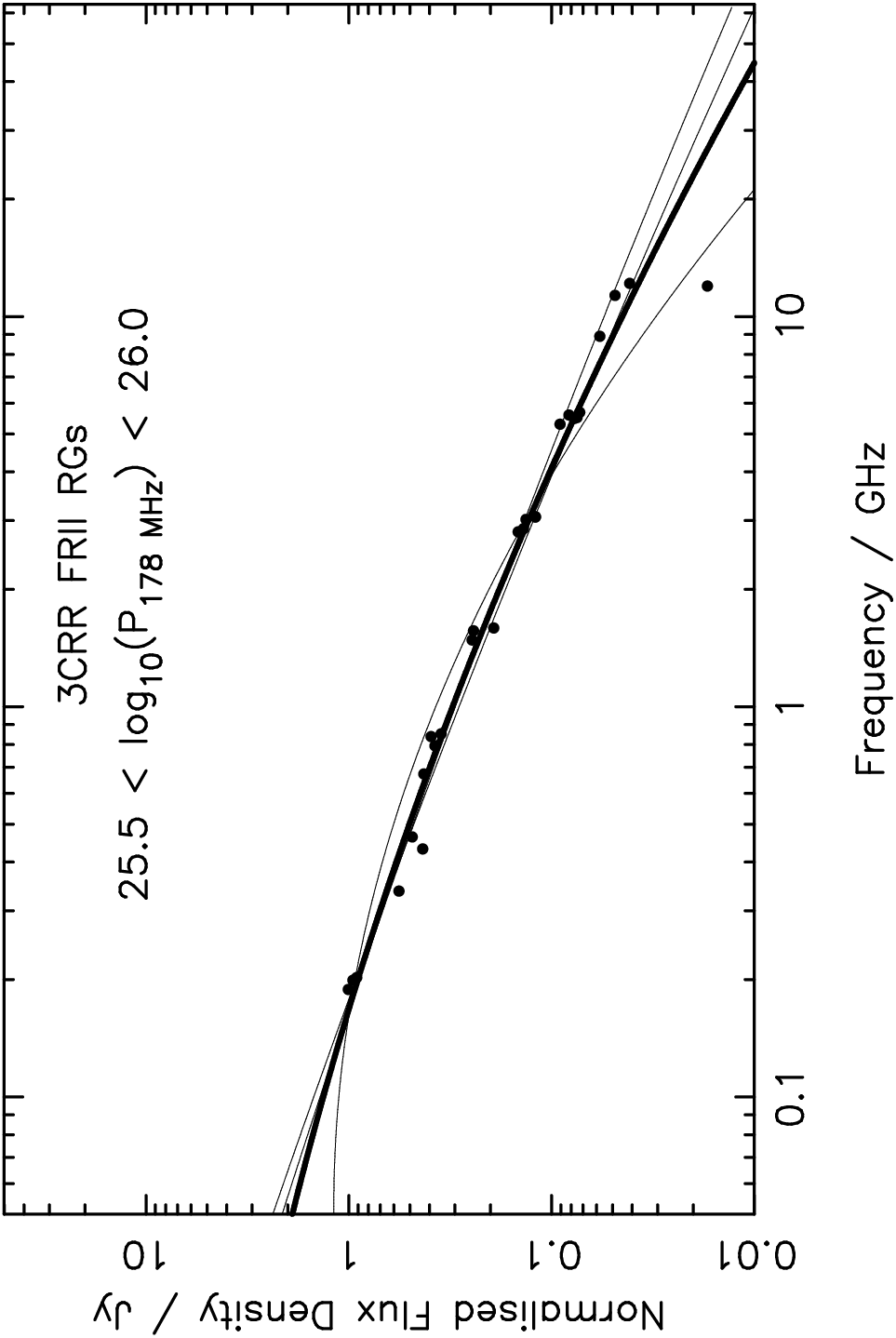}{3.0in}{270}{25}{35}{10}{500}
\plotfiddle{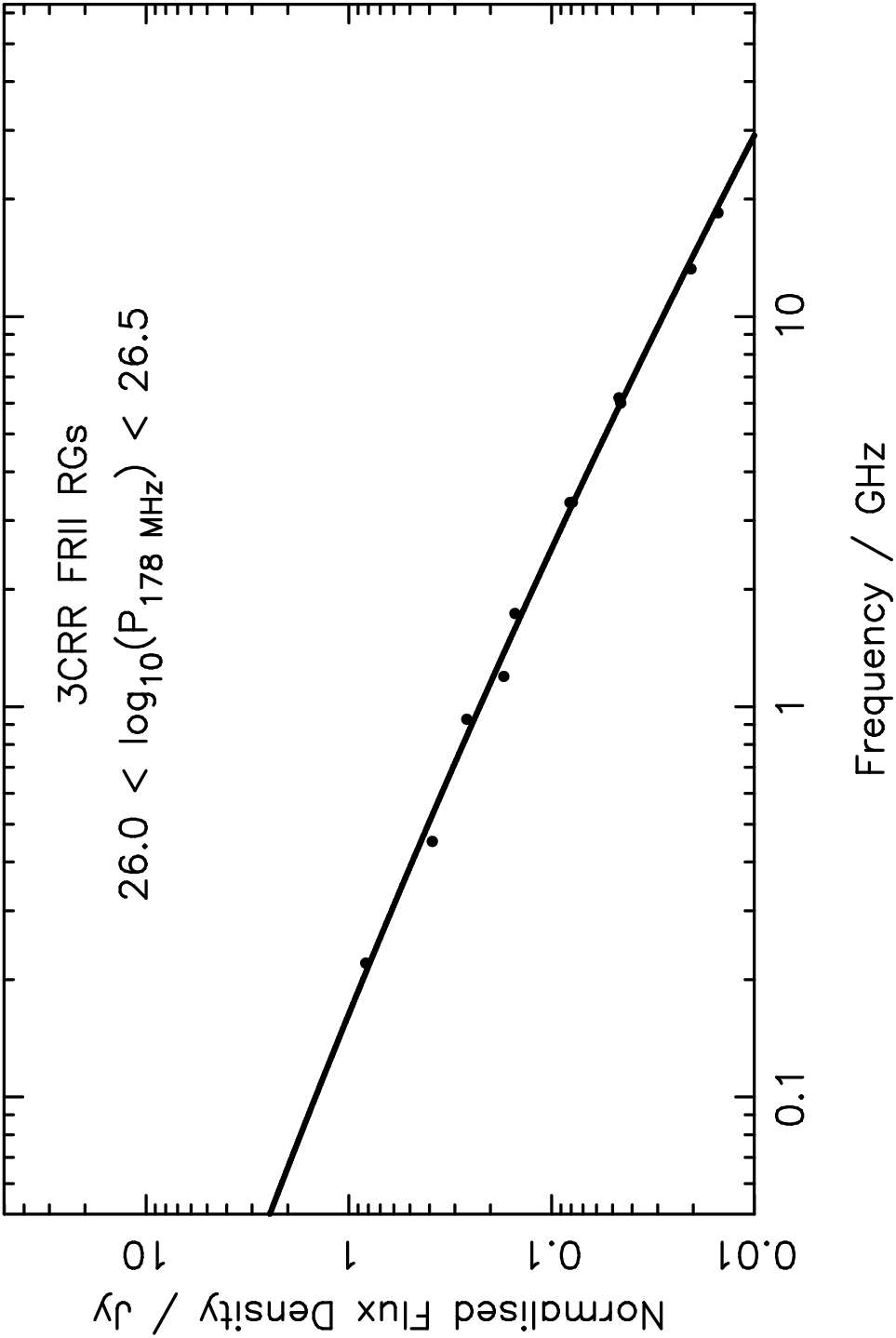}{3.0in}{270}{25}{35}{-180}{520}
\plotfiddle{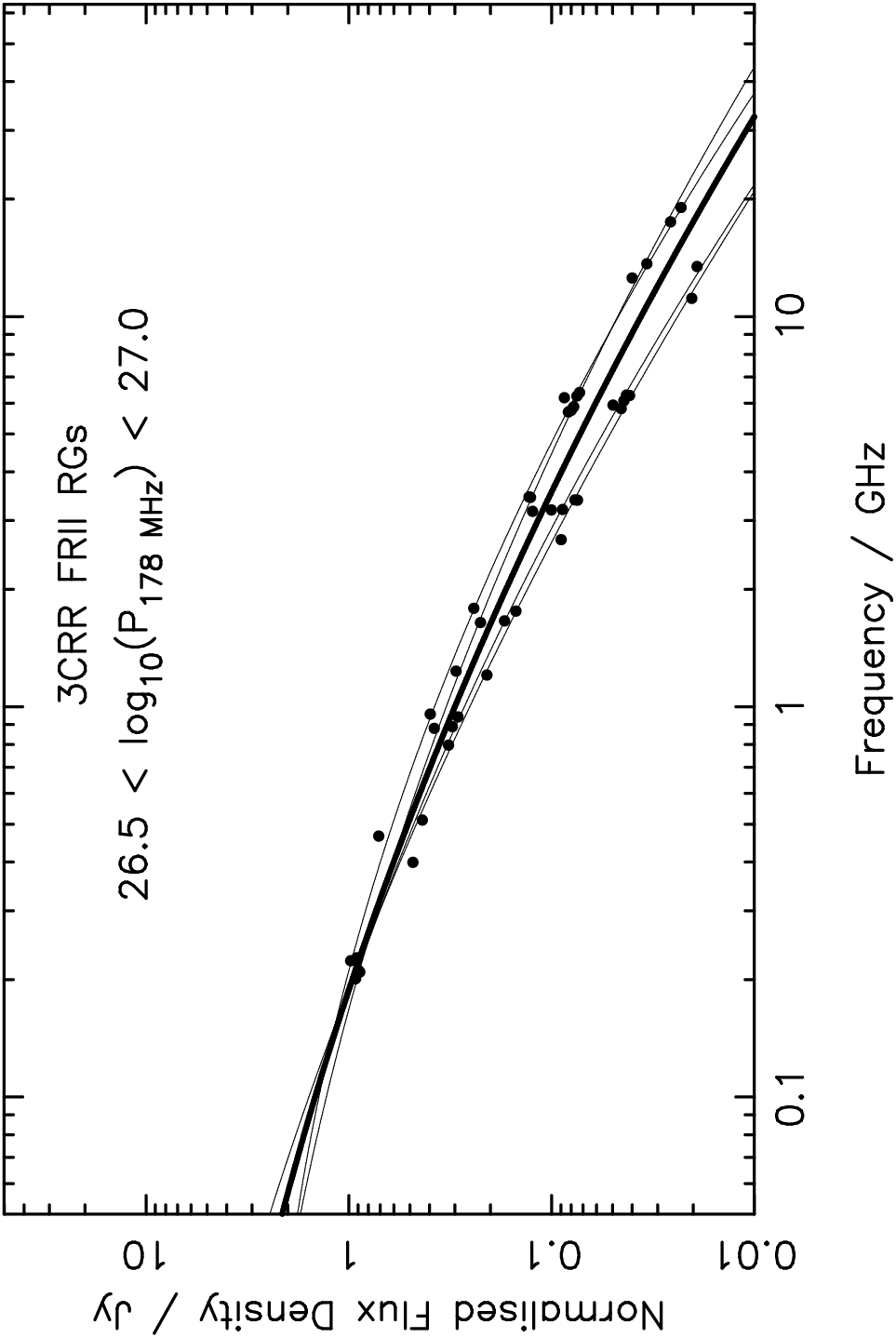}{3.0in}{270}{25}{35}{10}{750}
\vspace*{-7.5in}\caption{Fitted rest-frame
spectra for individual FRIIs (light lines) with radio powers
in the ranges show. The data are
($\bullet$) normalised to $S_{\fb}$ = 1 Jy. The
mean spectrum is the best-fit 
for $\nu_{rest} \ge$ 178 MHz (heavy line).}
\end{figure}

Because our evolution analysis (Jackson \& Wall 1999) uses
radio-source counts and luminosity-function estimates derived from
data at 151 MHz an 178 MHz, we considered only data at $\ge$ 178 MHz
(rest-frame) when deriving the best-fit for each galaxy.  We then
fitted the data on a radio-power range and FR-type basis, as shown in
Figures 2 and 3.  We fitted a quadratic; and found (Table 1)
increasing spectral curvature with increasing P (Laing \& Peacock
1980).

%\begin{figure}
%% from fitfig8.f (lost F77 code)
%\%plotfiddle{JacksonC_fig8.ps}{2.7in}{270}{35}{40}{-150}{240}
%\vspace*{-0.1in}\caption{Derived $\alpha$:$\log_{10}P$ functions
%for FRIs (dotted) and FRIIs (dashed).}
%\end{figure}

\begin{table}
\caption{Spectral fits by FR-type and radio power}
\vspace{0.1in}
\begin{tabular}{lrrccc}
      &                      & Number & \multicolumn{3}{c}{Quadratic
terms} \\ 
Class & $\log_{10}(P_{\fb})$ & of galaxies & A & 
B$(\log_{10}\nu)$ & C$(\log_{10}\nu)^{2}  $ \\
\hline
\\
FRI  & $24.0 \le P < 24.5$ & 5 & $-$0.4626 & $-$0.6424 & $-$0.0692 \\
     & $24.5 \le P < 25.0$ & 7 & $-$0.4563 & $-$0.7009 & $-$0.1132 \\
     & $25.0 \le P < 25.5$ & 2 & $-$0.6493 & $-$0.9213 & $-$0.0507 \\
     & $25.5 \le P < 26.0$ & 4 & $-$0.5582 & $-$0.8286 & $-$0.0279 \\
     & $26.0 \le P < 26.5$ & 3 & $-$0.5992 & $-$0.8794 & $-$0.0515 \\
% data & figures from sedfit1e to sedfit1a 
\\
FRII  & $25.0 \le P < 25.5$ & 2 & $-$0.5269 & $-$0.7128 & $-$0.0188 \\ 
     & $25.5 \le P < 26.0$ & 3 & $-$0.5114 & $-$0.7372 & $-$0.0982 \\
     & $26.0 \le P < 26.5$ & 1 & $-$0.6459 & $-$0.8567 & $-$0.0465 \\
     & $26.5 \le P < 27.0$ & 4 & $-$0.5228 & $-$0.8037 & $-$0.1145 \\
% data & figures from sedfit2a to sedfit2d
\hline
\end{tabular}
\end{table}

\begin{table}
\caption{Spectral fits by FR-type and radio power}
\vspace{0.1in}
\begin{tabular}{lrrccc}
      &                      & Number & \multicolumn{2}{c}{Quadratic
terms} \\ 
Class & $\log_{10}(P_{\fb})$ & of galaxies & 
B$(\log_{10}\nu)$ & C$(\log_{10}\nu)^{2}  $ \\
\hline
\\
FRI  & $24.0 \le P < 25.0$ & 12 & $-$0.6753 & $-$0.0901 \\
     & $25.5 \le P < 26.5$ &  7 & $-$0.8484 & $-$0.0395 \\
\\
FRII & $25.0 \le P < 26.0$ & 5 & $-$0.7249 & $-$0.0456 \\ 
     & $26.0 \le P < 27.0$ & 5  & $-$0.8136 & $-$0.1005 \\
% data & figures from sedfit2x & sedfit2y
\hline
\end{tabular}
\end{table}

We will adopt the fitted spectra of Table 2 in subsequent modelling of
radio source evolution. This will account for the spectral curvature
of the FRI and FRII (parent) radio source populations.

\end{document}